\documentclass[a4paper]{jpconf}
\usepackage{graphicx}
\begin{document}
\title{Nuclear Matter and Nuclear Dynamics}

\author{M Colonna}

\address{Laboratori Nazionali del Sud, INFN, via Santa Sofia 62, 
I-95123, Catania, Italy}

\ead{colonna@lns.infn.it}


\vskip -1.0cm
\section{Introduction}

The Equation of State (EOS) of nuclear matter plays a fundamental role
in the understanding of many aspects of nuclear physics and astrophysics. 
Transient states of nuclear matter far from normal conditions can be created
in terrestrial laboratories and 
many experimental and theoretical efforts have been devoted
to the study of nuclear reactions, from low to intermediate energies, as a 
possible tool to learn about the behavior of nuclear matter and its EOS.
In particular, the availability of exotic beams has opened
the way to explore, in laboratory conditions, new aspects of nuclear structure
and dynamics up to extreme ratios of neutron (N) to proton (Z) numbers.
Over the past years, 
measurements of isoscalar collective vibrations, collective flows and
meson production have contributed to constrain the EOS for symmetric matter
for densities up to five time the saturation value \cite{cons}. However, the EOS of
asymmetric matter has comparatively few experimental constraints: The
isovector part of the nuclear effective interaction (Iso-EOS) 
and the corresponding
symmetry energy are largely unknown 
far from normal density. 

The knowledge of the 
EOS  of asymmetric matter is very important at low densities, for studies
concerning neutron skins,
 pigmy resonances, nuclear structure at the drip lines, neutron star formation 
and crust, as well as at high densities, for investigations of 
 neutron star mass-radius relation, cooling, hybrid structure, transition
to a deconfined phase, formation of black holes. 
Hence
a large variety of phenomena, involving an enormous range of scales in
size, characteristic time and energy, but all based on nuclear processes
at fundamental level, are linked by the concept of EOS.   

From one side, this has stimulated new thorough studies of the ground state
energy of (asymmetric) nuclear systems, based on microscopic many-body
approaches, from which EOS and effective interactions can be extracted,  
that are essential for modeling heavy ion collisions (HIC)  
and the structure of neutron stars \cite{fuchswci,fantoni08,Baldo08,Lomb08_1}.   
Predictions of different many-body techniques will be compared in Section 2. 


On the other side, several observables which
are sensitive to the Iso-EOS and testable
experimentally, have been suggested
\cite{Isospin01,baranPR,WCI_betty,baoPR08}.
Taking advantage of new opportunities in 
theory (development of rather reliable microscopic transport codes 
for HIC)
 and in experiments (availability of very asymmetric radioactive beams, 
improved possibility of measuring event-by-event correlations), in the following Sections 3-5
we will review
new studies aiming to constrain the existing effective interaction 
models. We will discuss dissipative collisions in a wide range of beam 
energies, 
 from just above the Coulomb barrier up to the $AGeV$ range. 
Low to Fermi energies
 will bring information on the symmetry term around (below) normal density, 
while intermediate energies will probe high density regions.
As far as the high density behavior of asymmetric nuclear matter is concerned,
neutron stars, where densities ten times higher than the densities 
inside atomic nuclei may be reached in the core, appear as natural 
astrophysical laboratories that allow for a wide range of physical studies
linked to the properties of ultra-dense matter and of its EOS.
We will review recent developments in the field 
also in connection 
with the possibility that other degrees of freedom, besides the nucleonic ones,
(hyperons, quarks), may appear, see Section 6.
Finally, the link to the possible appearance of new phases of matter in charge asymmetric
HIC, already at intermediate energies, is discussed in Section 7.   


\section{Nuclear matter EOS and microscopic approaches}
The renewed interest in nuclear matter properties and EOS, that play
an important role in the understanding of nuclear structure and astrophysical 
phenomena, has led to the development of more detailed investigations of the 
ground state of nuclear many-body systems, where the energy functional is
evaluated starting from the bare nucleon-nucleon (NN) interaction. 
Therefore, from one side one is facing the problem of finding a fundamental 
scheme for the description of nuclear forces, valid from light nuclei up
to dense matter, that is still an open fundamental problem, and, on the other,
that of solving a complex many-body problem, that involves strongly 
spin-isospin dependent forces. 

It is well known that all non-relativistic many-body theories need at least
three-body forces (TBF) in the Hamiltonian. This is essential to reproduce
the correct saturation point of the nuclear EOS, extracted from 
phenomenological data \cite{Baldoreview}.
However, TBF cannot be constrained in a stringent way, due to the rather 
limited experimental data on three-nucleon systems. Phenomenological TBF
have been devised to reproduce the properties of light nuclei 
and to correct
the nuclear matter salturation point, leading to a new behavior of the whole
EOS of both symmetric and asymmetric nuclear matter \cite{Zhoureview}. 
However, a unique TBF, which is able to describe accurately light nuclei, 
as well as the nuclear matter saturation point, is not available yet. 
Hence, it is interesting to study to which extent the nuclear matter EOS, 
in a wide density range, is actually dependent on the choice of two-
and three-body forces, once the saturation point has been reproduced, i.e.
to test to what extent the reproduction of the saturation point constrains
the behavior of the whole EOS.
 
\begin{figure}[h]
\includegraphics[width=18pc]{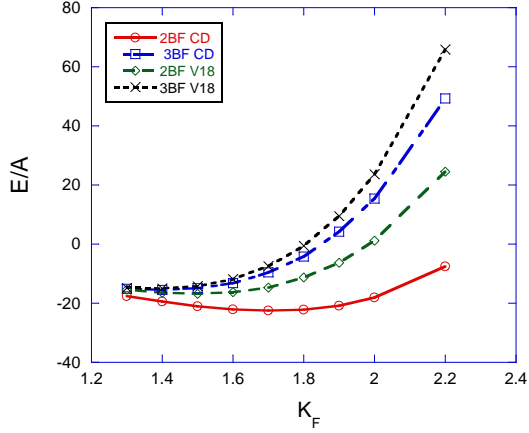}\hspace{2pc}%
\begin{minipage}[b]{14pc}\caption{\label{fig1_marcello}
Equation of state of symmetric nuclear matter for the CD Bonn and the v$_{18}$
two-body interactions and with the inclusion of three-body forces (TBF).
The energy per particle E/A is in MeV and the Fermi momentum $k_F$ is
in $fm^{-1}$. Taken from Ref.\cite{Baldo08}}
\end{minipage}
\end{figure}

In the following we will review recent  results obtained mostly in the context of the
many-body Bethe-Brueckner-Goldstone (BBG) theory \cite{Baldo08,Lomb08_1}, without the pretension to be exhaustive.    
Among the most accurate two-body forces, one has to consider the latest one 
of the Bonn potential series, that is more explicitely constructed from
meson exchange processes \cite{mac}.  
Also the Argonne $v_{18}$ NN interaction, that is
constructed by a set of two-body operators which arise naturally in
meson exchange processes, but with partially phenomenological form factors, 
can be considered as a very accurate one \cite{wiringa}. 
Concerning the phenomenological TBF (Urbana model), it contains a term that 
accounts for the so-called two-pion exchange contribution, with the creation of
an intermediate excited $\Delta$ state, and a phenomenological repulsive
part (Ropler excitation) \cite{panda}. 
As shown in Ref.\cite{Baldo08}, using exactly the same values of parameters
of the TBF  for the two choices of NN two-body interaction indicated above, the difference between
the corresponding EOS's, which are apparent in the case without TBF, are strongly reduced 
when TBF are introduced, not only around saturation, but for the whole density range. 
Hence, 
the overall effect of the same TBF on the EOS can be different according to the two-body
force adopted, being less repulsive in the case when the two-body force is less
attractive and pointing in the direction of making the EOS's closer to each other, see 
fig.\ref{fig1_marcello}.
This result is far from being trivial and it would be desirable to understand on physical
grounds what is the interplay of two- and three-body forces in determining the nuclear
EOS.

A different strategy is followed in Ref.\cite{Lomb08_1}: 
TBF can be evaluated on the basis
of a meson-exchange approach. Starting from two-body one-boson exchange (OBE) 
potentials, like
the Bonn or Nijmegen potentials (see Ref.\cite{Lomb08_1} and references therein), 
the meson exchange parameters of the TBF can be chosen
completely consistent with the given NN potential, i.e. the same parameters are used
in two and three-body forces. This kind of TBF involves the intermediate excitation of 
$\Delta$, Roper, and nucleon-antinucleon states by the exchange of $\pi,\rho,\sigma$, 
and $\omega$ mesons in the TBF diagrams. 
However, one must be aware of the fact that some parameters are not given by the two-body 
force, in particular the ones related to $\Delta$ and Roper excitation, and are evaluated
according to recent prescriptions \cite{Lomb08_1}.
Contrary to what is done in Ref.\cite{Baldo08},
within this kind of approach, the nuclear matter saturation point is not fitted 
with the help of a phenomenological TBF, but it naturally arises from the full calculation,
including two- and three-body forces and has to be checked against the experimental 
value.  
\begin{figure}[h]
\includegraphics[width=18pc]{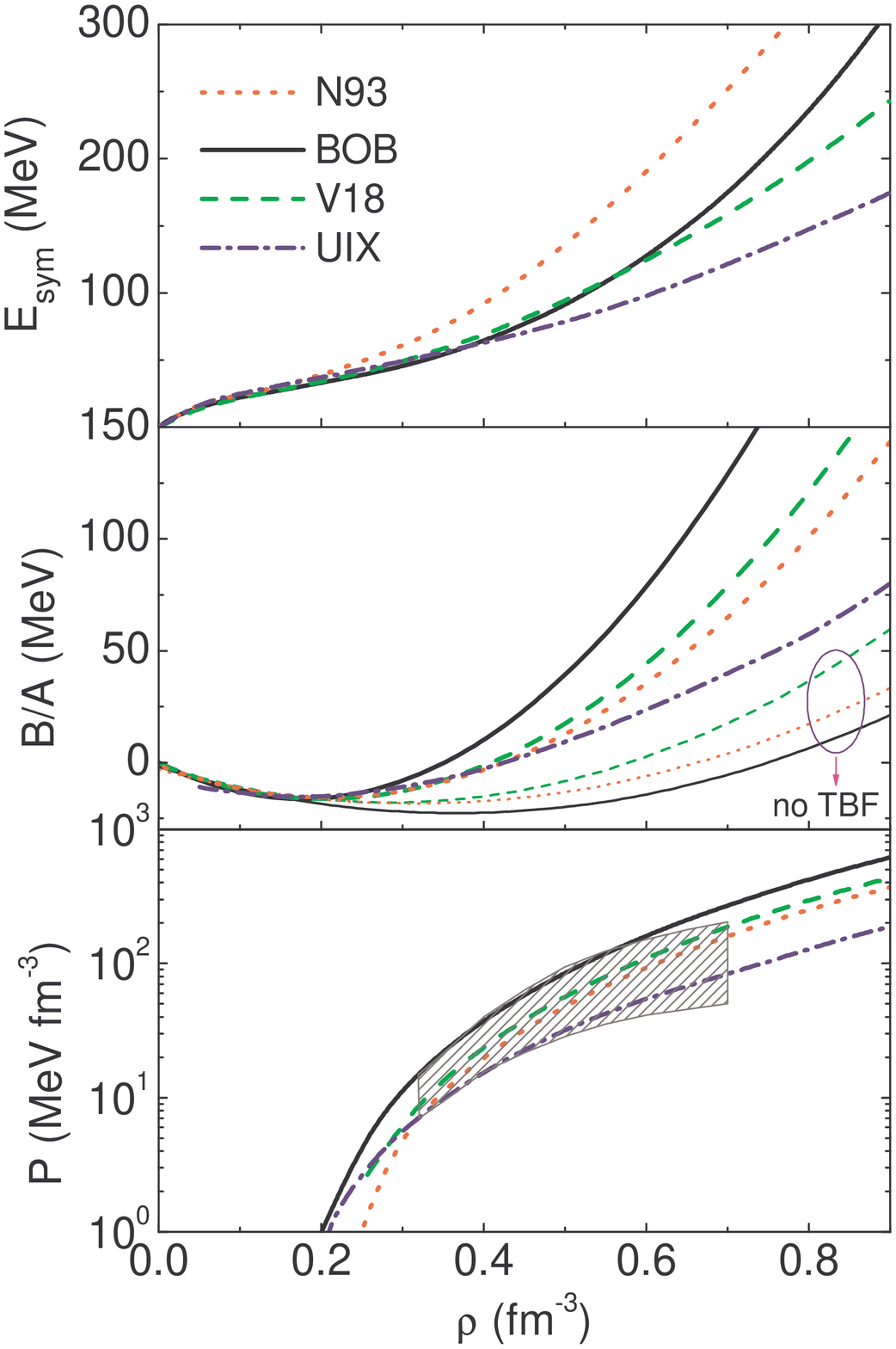}\hspace{2pc}%
\begin{minipage}[b]{14pc}\caption{\label{fig2_umb}
Symmetry energy (upper panel), binding energy per nucleon of symmetric
nuclear matter (central panel), and pressure of symmetric matter 
(lower panel), employing different interactions. The shaded region
indicates the constraints of Ref.\cite{cons}.
Taken from Ref.\cite{Lomb08}.}
\end{minipage}
\end{figure}
In fig.\ref{fig2_umb} (middle panel), we report the EOS obtained with only two-body forces and
with the full calculations, for different NN potentials.  Also in this case one can
see that the effect of the TBF is more repulsive when the two-body force is more attractive. 
This is the kind of compensation discussed before. 
The result corresponding to the Urbana (phenomenological)
TBF, coupled to the Argonne two-body interaction, is also shown on 
the figure (dot-dashed curve).  
Comparing dashed
and dot-dashed lines, it appears that a very different behavior is obtained for the
high density EOS and for the symmetry energy (middle panel) when using different TBF's, 
inspite of the fact that the saturation
point is reasonably reproduced in all cases. 
One may notice, in particular, the rather stiff high density behavior
obtained with the Bonn B (BOB) potential, for which the reproduction
of the saturation point properties is excellent. 
The corresponding symmetry energy exhibits an interesting behavior, that is soft at low density and
rather stiff at high density (see top panel).

From the results discussed here, one may conclude that
the density behavior of the symmetric matter EOS and symmetry energy 
is still a rather open problem and it is thus very appealing to try to get hints  
from the phenomenology of nuclear reactions and nuclear
astrophysics.  
We will review this kind of studies in the following, with a 
special emphasis on the information one can get on the density dependence of the
symmetry energy.

\section{Nuclear reactions and symmetry energy}
Heavy ion reactions with neutron rich nuclei, from low 
up to relativistic energies, 
can be used to study
the properties of the symmetry term of the nuclear interaction
in a wide range of densities. 

Nuclear reactions are modeled by solving transport equations
based on 
mean field theories, with correlations included via hard nucleon-nucleon
elastic and inelastic collisions and via stochastic forces, selfconsistently
evaluated from the mean phase-space trajectory, see 
\cite{chomazPR,baranPR}. 
Stochasticity is 
essential in 
order to get distributions as well as to allow for the growth of dynamical 
instabilities. 

In the energy range up to a few hundred AMeV, the appropriate tool is the 
so-called Boltzmann-Langevin equation (BLE) \cite{chomazPR}:
\begin{equation}
{{df}\over{dt}} = {{\partial f}\over{\partial t}} + \{f,H\} = I_{coll}[f] 
+ \delta I[f],
\end{equation}
 where $f({\bf r},{\bf p},t)$ is the one-body distribution function, 
or Wigner transform of the one-body density, 
$H({\bf r},{\bf p},t)$ the mean field Hamiltonian, 
$I_{coll}$ the two-body collision term 
incorporating the Fermi statistics of the particles,
and 
$\delta I[f]$ the fluctuating part of the
collision integral. 
For higher beam energies, a covariant formulation of transport equations will be considered, 
see Section 5.2. Hence effective interactions and 
the nuclear matter EOS can be considered as  ingredients of the transport codes
and from the comparison with experimental data one can finally get some hints
on nuclear matter properties. 

\begin{figure}[h]
\vskip 0.5cm
\includegraphics[width=16pc]{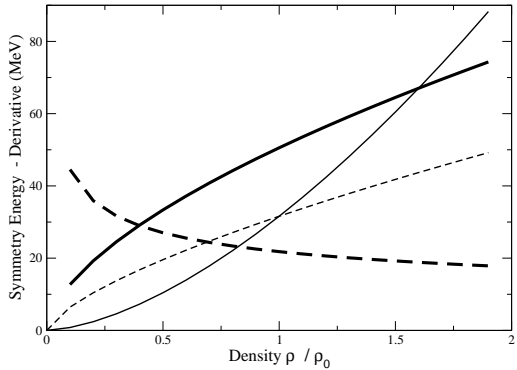}\hspace{2pc}%
\begin{minipage}[b]{16pc}\caption{\label{isoEOS}
Two representative effective parameterizations of the symmetry 
energy (thin lines): asystiff (full line) and asysoft (dashed line).
The tick lines show the corresponding derivatives.  
}
\end{minipage}
\end{figure}

We recall that 
the symmetry energy $E_{sym}$ appears in the energy density
$\epsilon(\rho,\rho_3) \equiv \epsilon(\rho)+\rho E_{sym} (\rho_3/\rho)^2
 + O(\rho_3/\rho)^4 +..$, expressed in terms of total ($\rho=\rho_p+\rho_n$)
and isospin ($\rho_3=\rho_p-\rho_n$) densities.
$E_{sym}$ gets a
kinetic contribution directly from basic Pauli correlations and a potential
part, $C(\rho)$,  from the highly controversial isospin dependence of 
the effective interactions: 
\begin{equation}
\frac{E_{sym}}{A}=\frac{E_{sym}}{A}(kin)+\frac{E_{sym}}{A}(pot)\equiv 
\frac{\epsilon_F}{3} + \frac{C(\rho)}{2\rho_0}\rho
\end{equation}
The sensitivity of the simulation results is tested against 
different choices of the density and momentum dependence of the
isovector part of the Equation of State. 
In the non-relativistic
frame, one can consider for 
the potential part of the symmetry energy, $C(\rho)$,
two opposite density
parametrizations (Iso-EOS) of the mean field \cite{bar02,baranPR}:
i) $\frac{C(\rho)}{\rho_0}=482-1638 \rho$ $(MeV fm^{3})$ for ``Asysoft'' 
EOS: ${E_{sym}/{A}}(pot)$
has a weak
density dependence close to the saturation, with an almost flat behavior below
the saturation density  $\rho_0$ and even decreasing at suprasaturation; ii) a constant coefficient, 
$C=32 MeV$, or a $C(\rho)$ coefficient linearly increasing with density, for the ``Asystiff'' EOS 
choice: the interaction part of the symmetry term displays
a faster decrease at lower densities and much stiffer above saturation, 
with respect to
the Asysoft case, see fig.\ref{isoEOS}.
The isoscalar section 
of the EOS is the same in both cases,
fixed requiring that  the saturation properties of symmetric nuclear matter
with a compressibility around $220MeV$  are reproduced.

The role  of the isospin degree of freedom on the dynamics of a nuclear reaction
can be discussed in a compact way by means of the 
chemical
potentials for protons and neutrons as a function of density $\rho$ and 
asymmetry 
$I = (N-Z)/A$ \cite{isotr05}. The $p/n$ currents can be expressed as
\begin{equation}
{\bf j}_{p/n} = D^{\rho}_{p/n}{\bf \nabla} \rho - D^{I}_{p/n}{\bf \nabla} I
\end{equation}
with $D^{\rho}_{p/n}$ the drift, and
$D^{I}_{p/n}$ the diffusion coefficients for transport, which are given 
explicitely
 in ref. \cite{isotr05}. Of interest for the study of isospin effects 
are the differences of currents 
between protons 
and neutrons which have a simple relation to the density dependence of the 
symmetry energy
\begin{eqnarray}
D^{\rho}_{n} - D^{\rho}_{p}  & \propto & 4 I \frac{\partial E_{sym}}
{\partial \rho} \,
 ,  \nonumber\\
D^{I}_{n} - D^{I}_{p} & \propto & 4 \rho E_{sym} \, .
\label{trcoeff}
\end{eqnarray}
Thus the isospin transport due to density gradients, i.e. isospin migration, 
depends on the slope of the symmetry energy, or the symmetry pressure, 
while the 
transport due to isospin concentration gradients, i.e. isospin diffusion, 
depends on
 the absolute value of the symmetry energy. 
Hence transport phenomena in nuclear reactions are directly linked to the 
EOS properties.

\vskip -1.0cm
\section{Symmetry energy at low density}
At sub-saturation density, the symmetry energy of nuclear matter seems to be 
under control from the theoretical point of view since different microscopic
many-body calculations agree among each other and TBF have a negligible
effects \cite{Baldoreview,Chiara}. According to these calculations, the symmetry energy
can be approximately described by $E_{sym}/A = 31.3 (\rho/\rho_0)^{0.6}$. 
The existing data and analyses extracted from nuclear reactions 
\cite{tsang92,Betty}
seem to point to a behavior of the type  
$E_{sym}/A \propto  (\rho/\rho_0)^\gamma$,
with $\gamma$ in the range $0.6-1$. 
Hence it is of great interest to pursue this investigation and devise new
observables to test the behavior of the symmetry energy at low density.

\subsection{The prompt dipole $\gamma$-ray emission in dissipative collisions}

The low-density behavior of the symmetry energy can be explored looking at dipole excitations
in dissipative charge asymmetric reactions around 10 AMeV.  
The possibility of an entrance channel bremsstrahlung dipole radiation
due to an initial different N/Z distribution was suggested at the beginning
of the nineties \cite{ChomazNPA563}. 
After several experimental evidences, in fusion as well as in deep-inelastic
reactions, \cite{PierrouPRC71,medea} and refs. therein,  
the process is 
now well understood and stimulating new perspectives
are coming from the use of radioactive beams.

During the charge equilibration process taking place
 in the first stages of dissipative reactions between colliding ions with
 different N/Z
ratios, a large amplitude dipole collective motion develops in the composite
dinuclear system, the so-called Dynamical Dipole mode. This collective dipole
gives rise to a prompt $\gamma $-ray emission which depends
on the absolute value of the initial amplitude, $D(t=0)$,
on the fusion/deep-inelastic dynamics and 
on the symmetry term, below saturation, that is acting as a restoring
force. Indeed this oscillation develops in the low density interface between
the two colliding ions (neck region). 

\begin{figure}
\begin{center}
\includegraphics*[angle=-90,scale=0.33]{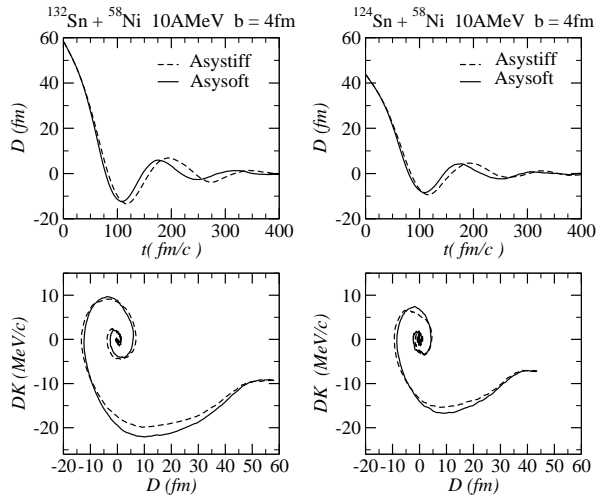}
\end{center}
\vskip -0.5cm
\caption{Dipole Dynamics at 10 AMeV, $b=4fm$ centrality. 
Left Panels: Exotic ``132'' system. Upper: Time evolution of dipole moment 
D(t) in real 
space; Lower: Dipole phase-space correlation (see text).
Right Panels: same as before for the stable ``124'' system.
Solid lines correspond to Asysoft EOS, the dashed to Asystiff EOS.}
\label{dip}
\end{figure}

A detailed description is obtained in mean field transport approaches
\cite{BaranPRL87}.
One can follow the time evolution
of the dipole moment
in the $r$-space,
 $D(t)= \frac{NZ}{A} ({R_{Z}}- {R_{N}})$ and in
$p-$space, $DK(t)=(\frac{P_{p}}{Z}-\frac{P_{n}}{N})$, 
being $R_p$, $P_{p}$
($R_n$, $P_{n}$) the centers of mass in coordinate and momentum space for protons (neutrons).
A nice "spiral-correlation"
clearly denotes the collective nature
 of the mode, see fig.\ref{dip}.
The ``prompt'' photon emission probability, with energy 
$E_{\gamma}= \hbar \omega$,
 can be estimated applying
a bremsstrahlung approach
 to the dipole evolution given from the BLE approach
\cite{BaranPRL87}:
\begin{equation}
\frac{dP}{dE_{\gamma}}= \frac{2 e^2}{3\pi \hbar c^3 E_{\gamma}}
 |D''(\omega)|^{2}  \label{brems},
\end{equation}
where $D''(\omega)$ is the Fourier transform of the dipole acceleration
$D''(t)$. We remark that in this way it is possible
to evaluate, in {\it absolute} values, the corresponding pre-equilibrium
photon emission.

We must add a
couple of comments of interest for the experimental selection of the Dynamical
Dipole: i) The centroid is always shifted to lower energies (large
deformation of the dinucleus); ii) A clear angular anisotropy should be present
since the prompt mode has a definite axis of oscillation
(on the reaction plane) at variance with the statistical giant dipole resonance ($GDR$).
These features have been observed in recent experiments \cite{medea}.



\begin{figure}
\begin{center}
\includegraphics*[scale=0.33]{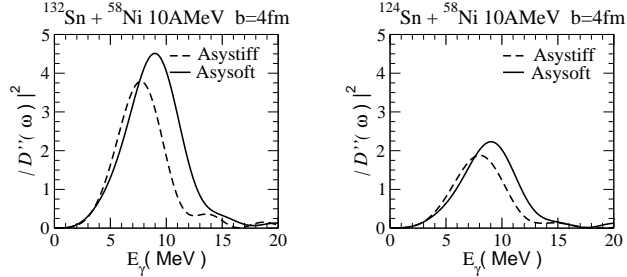}
\end{center}
\vskip -0.5cm
\caption{Left Panel, Exotic ``132'' system. Power spectra of the 
dipole acceleration at  $b=4$fm (in $c^2$ units).
Right Panel: Corresponding results for the stable ``124'' system.
Solid lines correspond to Asysoft EOS, the dashed to Asystiff EOS.}
\label{yield1}
\end{figure}

The use of unstable neutron rich projectiles would largely increase the
effect, due to the possibility of larger entrance channel asymmetries.
This can be observed in figs.\ref{dip},\ref{yield1}, where the results of
the reactions  $^{132}Sn+^{58}Ni$ and  $^{124}Sn+^{58}Ni$ are compared
\cite{dipang08}.

One can  notice in fig.\ref{dip} the large amplitude of the first oscillation for the ``132'' system, 
but also the delayed dynamics for the Asystiff EOS related to a weaker 
isovector restoring force. 
In fig.\ref{yield1} (Left Panel) we report the power spectrum, 
$\mid D''(\omega) \mid^2$ in semicentral
``132'' reactions, for different $Iso-EOS$ choices.
The gamma multiplicity is simply related to it, see Eq.(\ref{brems}).
The corresponding results for the stable ``124''  system are drawn
in the Right Panel.
As expected from the larger initial charge asymmetry, we clearly see an 
increase 
of the Prompt Dipole Emission for the exotic
n-rich beam. Such entrance channel effect allows also for 
a better observation of the Iso-EOS dependence. 
We remind that
in the Asystiff case we have a weaker restoring force 
for the dynamical dipole
in the dilute ``neck'' region, where the symmetry energy is smaller, see fig.3.
This is reflected in the 
lower value of the centroid, $\omega_0$,  as well as in the reduced total yield, as shown 
in fig.\ref{yield1}. The sensitivity of $\omega_0$ to the stiffness of the symmetry
energy is amplified by the increase of $D(t_0)$ using exotic,
more asymmetric beams. 

The prompt dipole radiation angular distribution is the result of the 
interplay between the collective oscillation life-time and the dinuclear 
rotation. In this sense one expects also a sensitivity to the $Iso-EOS$ of the
anisotropy, in particular for high spin event selections \cite{dipang08}.




\subsection{Isospin equilibration at the Fermi energies}

In this energy range the doorway state mechanism of the Dynamical Dipole
will disappear and so one can study a direct isospin transport in binary events.
This process also involves the low density neck region and is sensitive to
the low density behavior of $E_{sym}$, see Refs.\cite{tsang92},\cite{isotr07} and ref.s therein.  


 It is interesting to look at the asymmetries of the various parts 
of the reaction system in the exit channel:
emitted particles,  projectile-like (PLF) 
and target-like fragments (TLF), and in  the  case of ternary events,  intermediate mass
fragments (IMF).
 In particular, one can  study  the
 so-called Imbalance Ratio, which is defined as
\begin{equation}
R^x_{P,T} = \frac{2(x^M-x^{eq})}{(x^H-x^L)}~,
\label{imb_rat}
\end{equation}
with $x^{eq}=\frac{1}{2}(x^H+x^L)$.
 Here, $x$ is an isospin sensitive quantity
that has to be investigated with respect to
equilibration.   We consider primarily the asymmetry 
$\beta= I = (N-Z)/A$,
but also other quantities, such as isoscaling coefficients, ratios of 
production of light
 fragments, etc, can be of interest \cite{WCI_betty}. 
The indices $H$ and $L$ refer to the symmetric reaction
between the
heavy  ($n$-rich) and the light ($n$-poor)  systems, while $M$ refers to the
mixed reaction.
$P,T$ denote the rapidity region, in which this quantity is measured, in
particular the
PLF and TLF rapidity regions. Clearly, this ratio is $\pm1$ in
the projectile
and target regions, respectively, for complete transparency, and oppositely
for complete
rebound, while it is zero for complete equilibration.
\begin{figure}[h]
\includegraphics[width=20pc]{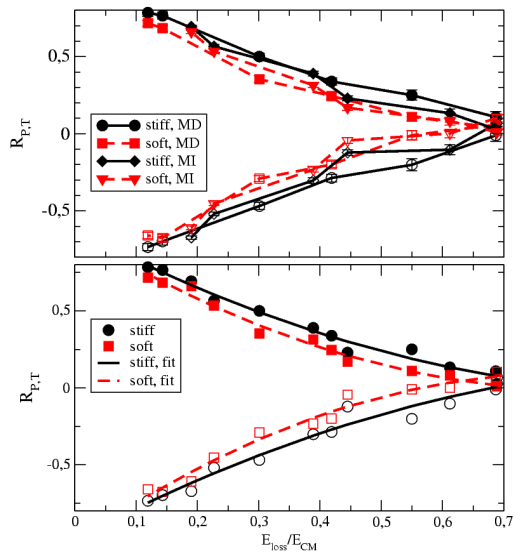}\hspace{2pc}%
\begin{minipage}[b]{14pc}\caption{\label{imb_eloss}
Imbalance ratios as a function of relative energy loss. 
Upper: Separately for 
stiff (solid) and soft (dashed) iso-EOS, and for 
two parameterizations of the isoscalar part of the interaction: 
MD 
(circles and squares) and MI 
(diamonds and triangles), in the projectile region (full symbols)
 and the target region 
(open symbols).
Lower: Quadratic fit to all points for the stiff (solid), resp.
 soft (dashed) 
iso-EOS.}
\end{minipage}
\end{figure}  


In a simple model one can show that the imbalance ratio mainly depends on two
quantities: the strength of the symmetry energy and the interaction
time between the two reaction partners.
Let us take, for instance, the asymmetry $\beta$ of the PLF (or TLF) as the
quantity $x$.
At a first order approximation, in the mixed reaction this quantity relaxes
towards
its complete equilibration value, $\beta_{eq} = (\beta_H + \beta_L)/2$, as
\begin{equation}
\label{dif_new}
\beta^M_{P,T} = \beta^{eq} + (\beta^{H,L} -  \beta^{eq})~e^{-t/\tau},
\end{equation}
where $t$ is the time elapsed while the reaction partners stay in contact
(interaction time) and the damping $\tau$ is mainly connected to the strength 
of the symmetry energy \cite{isotr07}. 
Inserting this expression into Eq.(\ref{imb_rat}), one obtains
$ R^{\beta}_{P,T} = \pm e^{-t/\tau}$ for the PLF and TLF regions, respectively.
Hence the imbalance ratio can be considered as a good observable to 
trace back the strength
of the symmetry energy from the reaction dynamics
provided a suitable selection of the interaction time is performed.
 


The centrality dependence of the Imbalance Ratio, for (Sn,Sn) collisions at 35 and 50 AMeV,
has been investigated in experiments as well as in theory
\cite{isotr05,tsang92}. We report here a new analysis
 which appears experimentally more selective \cite{isotr07}. 
Longer interaction times should be correlated to
a larger 
dissipation. It is then natural to look at the correlation between
the imbalance ratio and the total kinetic energy loss.
In this way one can also better disentangle dynamical effects of the isoscalar 
and isovector part of the EOS, see \cite{isotr07}.

It is seen in fig.\ref{imb_eloss} (top) that the curves for the 
{\it Asysoft} EOS (dashed) are 
generally lower in the projectile region
 (and oppositely for the target region), i.e. show 
more equilibration, than those for the {\it Asystiff} EOS, due to the higher value
of the symmetry energy at low density. 
To emphasize 
this trend, all  the values for the
stiff (circles) and 
the soft (squares) iso-EOS, corresponding to different impact
parameters, beam energies and also to two possible parameterizations of the isoscalar part of the
nuclear interaction (MD and MI) are collected together in the bottom part of the figure. 
One can see that all the points essentially follow a given line,
depending only on the symmetry energy parameterization adopted.  
It is seen,
 that there is a systematic effect of the symmetry energy of the order 
of about 20 percent, 
which should be measurable. The correlation suggested in fig.\ref{imb_eloss}
should represent 
a general feature of isospin diffusion, and it would be of great 
interest to verify  it experimentally.


\subsection{Dipole emission in peripheral collisions at relativistic energies}
Here we briefly review results obtained with more
mascroscopic models on dipole excitation in peripheral
collisions, such as  $^{40}$Ca + $^{24}$O at 500 AMeV. 
In Ref.\cite{Edo} a relativistic treatment is elaborated, to derive the 
equations of motion of (projectile and target) neutron and proton centroids, under the action 
of the restoring isovector nuclear force, taking into account in the excitation of the dipole
oscillations, apart from Coulomb effects,  
also the contribution of the nuclear potential, treated with a Wood-Saxon potential well. 
It is observed that mean-field effects give a strong contribution to the excitation of 
the dipole oscillation, that is connected to the neutron skin of the target nucleus,
leading to a different attraction between the target protons or neutrons and
the projectile nucleons.  
The actual cross section for one- or two-phonon excitations is then evaluated according to the
expression $\sigma_\alpha = 2\pi \int_0^\infty P_\alpha(b)T(b)bdb$,
where $P_\alpha(b)$ is the probability, at the impact paramter $b$,  
for a given reaction channel $\alpha$. This is related to the number of phonons, 
$N_{ph} = E_\infty/(\hbar\omega)$ excited by the driving force, being $E_\infty$ the energy
associated with the dipole oscillations at asymptotic distance. 
$T(b)$ are attenuation factors that take into account the 
depopulation of the reaction channels due to couplings not explicitely included in the truncated 
working space.   

Comparing the obtained cross sections to experimental data,  from these studies
one can get information on the neutron skin structure 
of neutron-rich systems 
that, in turn, depends on the
low density behavior of the symmetry energy.

\vskip -1.0cm
\subsection{Isospin distillation in central collisions}

In central collisions at 30-50 AMeV, where the full disassembly of the system
into many fragments is observed, one can study specifically properties of
liquid-gas phase transitions occurring in asymmetric matter 
\cite{bao197,chomazPR,baranPR,baoPR08,Aldo}. 
For instance,
in neutron-rich matter, phase co-existence leads to
a different asymmetry
in the liquid and gaseous phase:  fragments (liquid) appear more symmetric
with respect to the initial matter, while light particles (gas) are
more neutron-rich.
\begin{figure}[t]
\centering
\includegraphics[width=6.0cm]{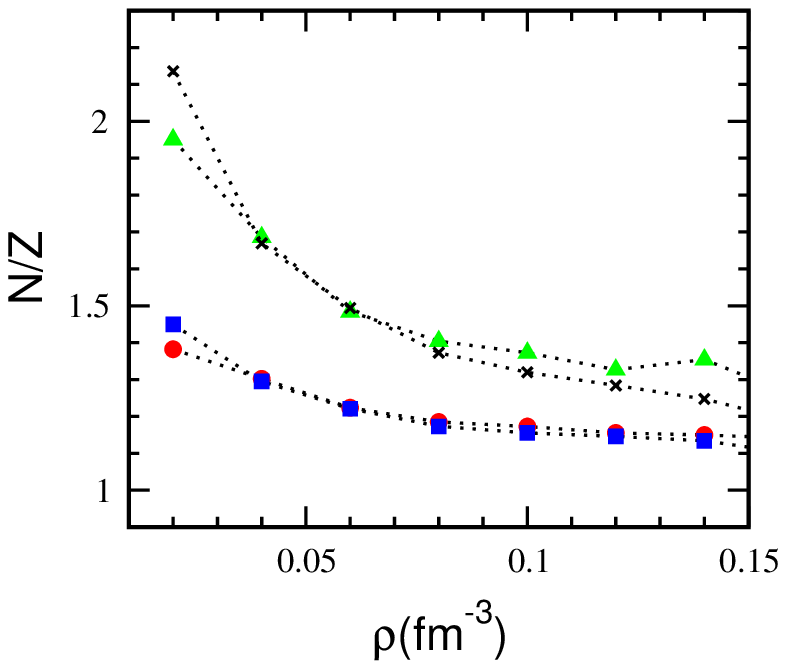}
\hskip 0.5cm
\includegraphics[width=6.0cm]{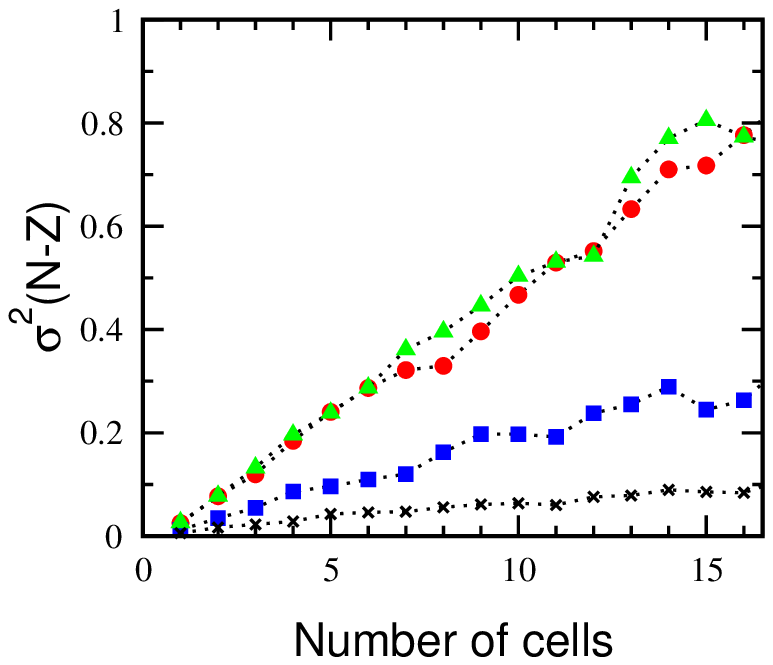}
\caption{Left Panel: Correlation between the N/Z and the density of the
domains, obtained with the two asy-EOS in the decomposition of unstable nuclear matter,
with asymmetry $\beta = 0.1$ (squares and circles) and $\beta = 0.2$
(triangles and stars). Right: Variance of the (N-Z) content of the density
domains as a function of their volume \cite{Mat08}. }
\label{ciccio}
\end{figure}
The amplitude of this "distillation" effect, that is connected to the presence of density
gradients in the system, 
depends on
 specific properties of the isovector part of the nuclear interaction,
namely on the value and the derivative of the symmetry
energy at low density (see also the discussion in Section 3).
Indeed, since the symmetry energy
increases with density,  
it is energetically more convenient for asymmetric systems to transfer the neutron
excess to the gas phase, while fragments (liquid phase) become more symmetric. 
These features can be easily evidenced by studying fragment formation in unstable  
asymmetric nuclear matter inside a box.
In fig.\ref{ciccio} (left) the N/Z of the domains formed inside the box is reported
as a function of the density, for systems with two initial charge asymmetry values and for
two Iso-EOS employed. The variance of the (N-Z) difference inside the 
density domains is displayed on the right part of the figure, as a function of 
the volume of the domain considered. One can see that, both the average N/Z and
the isovector variance decrease as the density increases,  leading to a more
symmetric and less fluctuating liquid phase. However, fluctuations are rather large
in the Asysoft case (circles and triangles) \cite{Mat08}. 
    

The investigation of isospin distillation is interesting in a more general context:
In heavy ion collisions the dilute phase appears during the expansion
of the interacting matter.
Thus we study effects of the coupling of expansion, fragmentation and 
distillation in a 
two-component (neutron-proton) system \cite{col07}.
Let us focus now on actual collisions, considering symmetric 
reactions
between systems having three different initial asymmetry:
$^{112}Sn + ^{112}Sn,^{124}Sn + ^{124}Sn,
^{132}Sn + ^{132}Sn,$ with $(N/Z)_{in}$ = 1.24,1.48,1.64, respectively.
The considered
beam energy is 50 AMeV.
\begin{figure}[t]
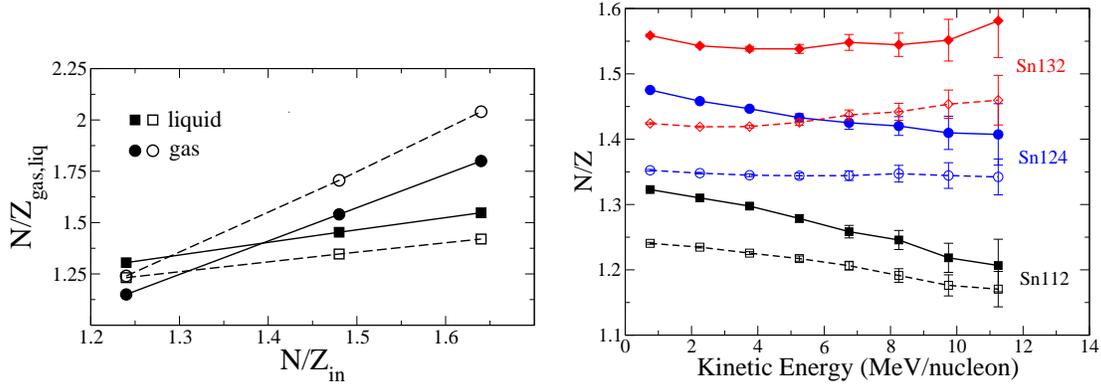

\vskip 1.cm
\centering
\includegraphics[width=7.0cm]{fig9.eps}
\hskip 0.5cm
\includegraphics[width=7.0cm]{fig10.eps}
\caption{Left Panel.The N/Z of the liquid (squares) and of the gas (circles) 
phase
is displayed as a function of the system initial N/Z.
Full lines and symbols refer to the asystiff parameterization. Dashed
lines and open symbols are for asysoft.
Right Panel.
The fragment N/Z as a function of the kinetic energy.
Full lines: Asystiff;  Dashed lines: Asysoft.
}
\label{iso_kin}
\end{figure}
The average N/Z of emitted nucleons (gas phase) and IMF's
is presented in fig.\ref{iso_kin} (Left Panel) as a function of the initial $(N/Z)_{in}$
of the three colliding systems.
One observes a clear Isospin-Distillation effect, i.e.
the gas phase (circles) is more neutron-rich than the IMF's (squares).
This is particularly evident in the
Asysoft case
due to the larger value of the symmetry energy  and of its derivative at low density
\cite{baranPR}.


Fragmentation originates from the break-up of a composite source
that expands with a given velocity field.
Since neutrons and protons experience different forces,
one may expect a different radial flow for the two species.
In this case,  the N/Z composition of the source would not be uniform,
but would depend on the radial distance from the center or mass or,
equivalently, on the local velocity.
It has been recently proposed that
this trend should then be reflected in a clear correlation between
isospin content and kinetic energy of the formed IMF's \cite{col07}.

This observable is plotted in fig.\ref{iso_kin} (Right Panel) for the three reactions.
The behaviour observed is rather sensitive to
the Iso-EOS.
For the proton-rich system, the N/Z decreases with the
fragment kinetic energy, expecially in the Asystiff case (left), where the
symmetry energy is relatively small at low density.
In this case, the Coulomb repulsion
pushes the protons towards the surface of the system. Hence, more
symmetric fragments acquire larger velocity.
The decreasing trend is less pronounced in the Asysoft case
(right) because Coulomb effects on protons
are counterbalanced by the larger
attraction of the symmetry potential at low density.
In systems with higher initial asymmetry, the decreasing
trend is inversed, due to the larger neutron repulsion in neutron-rich
systems. 
This analysis reveals the existence of
significant, EOS-dependent correlations between
the $N/Z$ and the kinetic energy of
IMF's produced in central collisions.
This appears as a promising experimental observable to be investigated, though
fragment secondary effects are expected to reduce the sensitivity to the
iso-EOS \cite {col07}.

\begin{figure}
\centering
\vskip -0.5cm
 \includegraphics[width=18pc]{fig11.eps}
 \includegraphics[width=18pc]{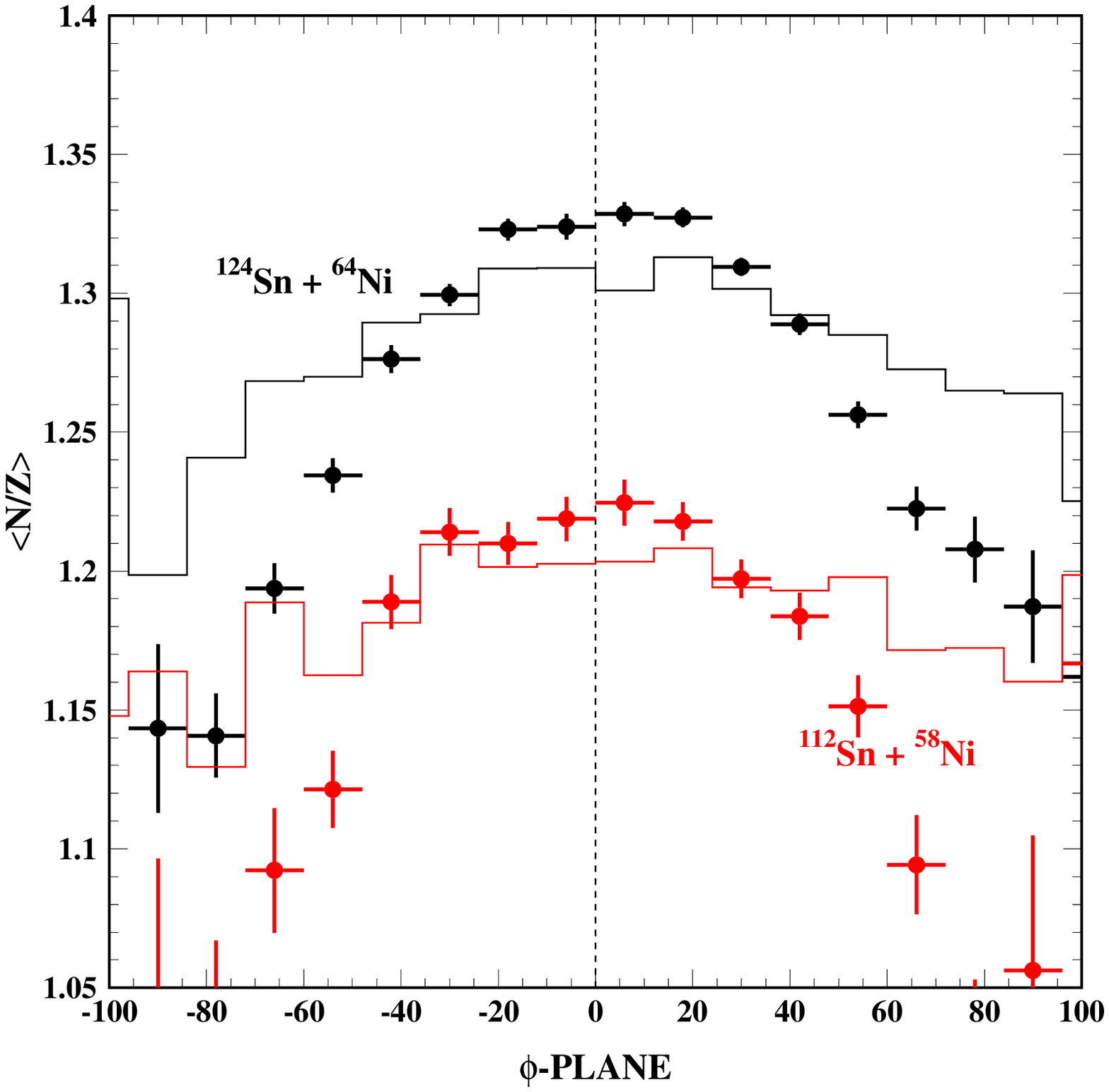}
\caption{
Left: Asymmetry of IMF's (circles) and PLF-TLF (squares), as a function of the
system initial asymmetry, for two iso-EOS choices: Asystiff (full lines) and
Asysoft (dashed lines).  
Right: Exp. results on 
correlation between $N/Z$ of $IMF$ and $alignement$ in ternary
events of  the $^{124}Sn+^{64}Ni$ and  $^{112}Sn+^{58}Ni$
reactions at $35~AMeV$. Points correspond to fast 
formed $IMF$s; histograms to all $IMF$s at 
mid-rapidity (including statistical emissions). From Ref.\cite{defilposter}}.
\label{nzphi}
\end{figure}

\vskip -1.0cm
\subsection{Isospin dynamics in neck fragmentation at Fermi energies}

It is now quite well established that the largest part of the reaction
cross section for dissipative collisions at Fermi energies goes
through the {\it Neck Fragmentation} channel, with $IMF$s directly
produced in the interacting zone in semiperipheral collisions on very short
time scales \cite{wcineck}. It is possible to predict interesting 
isospin transport 
effects for this 
fragmentation mechanism since clusters are formed still in a dilute
asymmetric matter but always in contact with the regions of the
projectile-like and target-like remnants almost at normal densities.
As discussed in Sect.3, in presence of density gradients the isospin transport
is mainly ruled by drift coefficients and so
we expect a larger neutron flow to
 the neck clusters for a stiffer symmetry energy around saturation  
\cite{baranPR}.
This is shown in fig.\ref{nzphi} (left), where the asymmetry of the neck region 
of charge asymmetric reactions is plotted for two Iso-EOS choices. 

A very nice new analysis has been performed on the $Sn+Ni$ data at $35~AMeV$
by the Chimera Collab.\cite{defilposter},
 see fig.\ref{nzphi} right panel.
A strong correlation between neutron enrichement and fragment alignement (when the 
short emission time selection is enforced) is seen, that points to 
a stiff behavior of the symmetry energy, for which a large neutron
enrichment of neck fragments is seen (left). 
This represents a 
clear evidence in favor of a relatively large slope (symmetry pressure) 
around saturation. We note a recent confirmation from structure data,
i.e. from monopole resonances in Sn-isotopes \cite{garg_prl07}.

\vskip -1.0cm
\section{Iso-EOS at supra-saturation density} 
\subsection{Effective mass splitting and collective flows}


The problem of Momentum Dependence in the Isovector
channel ($Iso-MD$) of the nuclear interaction 
is still very controversial and it would be extremely
important to get more definite experimental information,
see the recent refs. 
\cite{BaoNPA735,rizzoPRC72}. 
Exotic Beams at intermediate energies (100-500 AMeV  ) are
of interest in order to have high momentum particles and to test regions
of high baryon (isoscalar) and isospin (isovector) density during the
reaction dynamics.
Transport codes are usually implemented with 
different $(n,p)$ momentum dependences, see 
\cite{rizzoPRC72,BaoNPA735}. 
This  allows one to follow the dynamical
effect of opposite n/p effective mass ($m^*$) splitting while keeping the
same density dependence of the symmetry energy \cite{isotr07}.

For central 
collisions in the interacting zone baryon densities about
$1.7-1.8 \rho_0$ can be reached in a transient time of the order of 15-20 fm/c. 
The system 
is quickly expanding and the Freeze-Out time is around 50fm/c.
Here it is very interesting to study again the collective response of the system.
Collective flows are very good candidates since they are 
expected to be 
rather sensitive to the momentum
dependence of the mean field, see \cite{DanielNPA673,baranPR}.
The transverse flow, 
$V_1(y,p_t)=\langle \frac{p_x}{p_t} \rangle$,
provides information on the anisotropy of 
nucleon emission on the reaction plane.
Very important for the reaction dynamics is the elliptic
flow,
$V_2(y,p_t)=\langle \frac{p_x^2-p_y^2}{p_t^2} \rangle$.
 The sign of $V_2$ indicates the azimuthal anisotropy of emission:
on the reaction
plane ($V_2>0$) or out-of-plane ($squeeze-out,~V_2<0$)
\cite{DanielNPA673}.
The $Iso-MD$ of the fields can be tested 
just evaluating the difference of neutron/proton transverse and elliptic 
flows 
$
V^{(n-p)}_{1,2} (y,p_t) \equiv V^n_{1,2}(y,p_t) - V^p_{1,2}(y,p_t)
$ 
at various rapidities and transverse momenta 
in semicentral ($b/b_{max}=0.5$) collisions, such as
 $^{197}Au+^{197}Au$ at $400~AMeV$.
For the nucleon elliptic flows
 the mass splitting
effect is evident at all rapidities, and nicely increasing at larger
rapidities and transverse momenta, with more neutron flow when
$m_n^*<m_p^*$.
\begin{figure}[h]
\includegraphics[width=16pc]{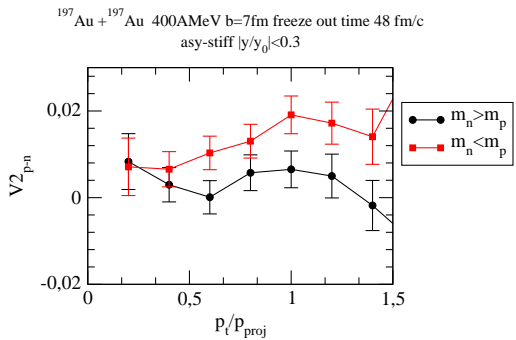}\hspace{2pc}%
\begin{minipage}[b]{14pc}\caption{\label{v2dif}
Transverse momentum dependence of the difference between proton 
and neutron $V_2$ flows, at mid-rapidity, in a 
semi-central
reaction Au+Au at 400AMeV. Taken from Refs.\cite{vale08},\cite{Erice08}.
}
\end{minipage}
\end{figure}  
From fig.\ref{v2dif} it is clear how at, mid-rapidity, the mass splitting effects 
are more evident
for higher tranverse momentum selections, i.e. for high
density sources. In particular
the elliptic flow difference becomes negative when
$m_n^*<m_p^*$, revealing a faster neutron emission and so more neutron
squeeze out (more spectator shadowing).
In correspondance the proton flow
is more negative (more proton squeeeze out) when $m_n^*>m_p^*$.
Due to the difficulties in
measuring neutrons, one could investigate the difference between
light isobar flows, like $^3H$ vs. $^3He$.
We still expect to see effective mass splitting effects. 

\subsection{
Meson production in relativistic heavy ion
collisions}
The phenomenology of isospin effects on heavy ion reactions
at relativistic energies (few $AGeV$ range) is extremely rich and can allow
a ``direct'' study of the covariant structure of the isovector interaction
in a high density hadron medium. 
In this energy range, one has to work within a relativistic transport frame,
beyond the cascade picture, 
 consistently derived from effective Lagrangians, where isospin effects
are accounted for in the mean field and collision terms.
Heavy ion collisions are described solving
the covariant transport equation of the Boltzmann type (RBUU)
\cite{FuchsNPA589},
and applying
a Monte-Carlo procedure for the hard hadron collisions.
The collision term includes elastic and inelastic processes involving
the production/absorption of the $\Delta(1232 MeV)$ and $N^{*}(1440
MeV)$ resonances as well as their decays into pion channels
 \cite{ferini06}.

The effective Lagrangian approach to the hadron interacting system can be
extended to the isospin degree of freedom: within the same frame equilibrium
properties ($EOS$, \cite{qhd}) and transport dynamics 
can be consistently derived.
Within a covariant picture of the nuclear mean field, 
 for the description of the symmetry energy at saturation
(a) only the Lorentz vector $\rho$ mesonic field, 
and (b) both, the vector $\rho$ (repulsive) and  scalar 
$\delta$ (attractive) effective 
fields are included. 
In the latter case a rather intuitive form of the symmetry energy can be
obtained \cite{baranPR}
\begin{equation} 
E_{sym} = \frac{1}{6} \frac{k_{F}^{2}}{E_{F}} +  
\frac{1}{2} 
\left[ f_{\rho} - f_{\delta}\left( \frac{m^{*}}{E_{F}} \right)^{2} 
\right] \rho_{B}. 
\label{esym3} 
\quad . 
\end{equation} 
The competition between scalar and vector fields leads
to a stiffer symmetry term at high density \cite{baranPR}. 

To discuss meson production, 
the starting point is
a simple phenomenological version of the Non-Linear (with respect to the 
iso-scalar, Lorentz scalar $\sigma$ field) effective nucleon-boson 
field theory,  
the Quantum-Hadro-Dynamics \cite{qhd}. 
According to this picture 
the presence of the hadronic medium leads to effective masses and 
momenta $M^{*}=M+\Sigma_{s}$,   
 $k^{*\mu}=k^{\mu}-\Sigma^{\mu}$, with
$\Sigma_{s},~\Sigma^{\mu}$
 scalar and vector self-energies. 
For asymmetric matter the self-energies are different for protons and 
neutrons, depending on the isovector meson contributions. 
The 
corresponding models are named 
$NL\rho$ and $NL\rho\delta$, respectively, and
just $NL$ for the case without isovector interactions. 
For the more general $NL\rho\delta$ case  
the self-energies 
of protons and neutrons read:
\begin{eqnarray}
\Sigma_{s}(p,n) = - f_{\sigma}\sigma(\rho_{s}) \pm f_{\delta}\rho_{s3}, 
\nonumber \\
\Sigma^{\mu}(p,n) = f_{\omega}j^{\mu} \mp f_{\rho}j^{\mu}_{3},
\label{selfen}
\end{eqnarray}
(upper signs for neutrons), where $\rho_{s}=\rho_{sp}+\rho_{sn},~
j^{\alpha}=j^{\alpha}_{p}+j^{\alpha}_{n},\rho_{s3}=\rho_{sp}-\rho_{sn},
~j^{\alpha}_{3}=j^{\alpha}_{p}-j^{\alpha}_{n}$ are the total and 
isospin scalar 
densities and currents and $f_{\sigma,\omega,\rho,\delta}$  are the coupling 
constants of the various 
mesonic fields. 
$\sigma(\rho_{s})$ is the solution of the non linear 
equation for the $\sigma$ field \cite{baranPR}.
From the form of the scalar self-energies we note that in n-rich environment
the neutron effective masses are definitely below the proton ones.

Kaon production has been proven to be a reliable observable for the
high density $EOS$ in the isoscalar sector 
\cite{FuchsPPNP56}.
Here we show that the $K^{0,+}$
production (in particular the $K^0/K^+$ yield ratio) can be also used to 
probe the isovector part of the $EOS$
\cite{ferini06}.
Pion and kaon production are analyzed in central $^{197}Au+^{197}Au$ collisions in 
the $0.8-1.8~AGeV$
beam energy range \cite{ferini06}.  
\begin{figure}[t] 
\begin{center}
\includegraphics[scale=0.35]{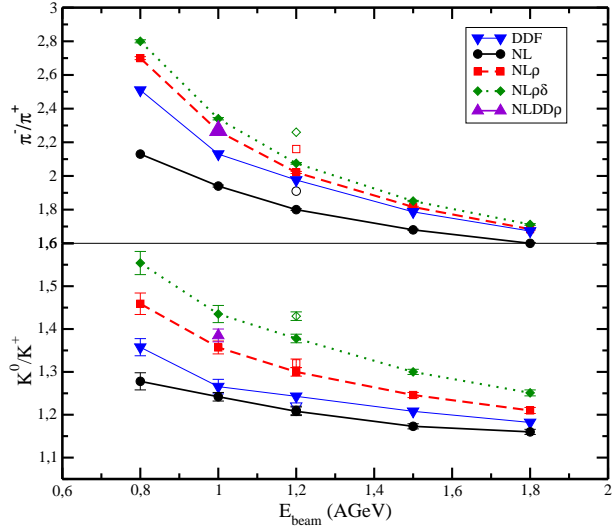}
\vskip -0.3cm
\caption{$\pi^{-}/\pi^+$ (top) and $K^{0}/K^+$ (bottom)
ratios as a function of the incident energy for central Au + Au collisions,
for several parameterizations of the isovector nuclear interaction. 
The open symbols at 1.2 AGeV show the corresponding results for a 
$^{132}$Sn + $^{124}$Sn collision, more neutron rich.   

}
\label{kaon1} 
\end{center}
\end{figure} 


When isovector fields are included the symmetry potential energy in 
neutron-rich matter is repulsive for neutrons and attractive for protons.
In a heavy ion collision this leads to a fast, pre-equilibrium, emission of neutrons.
 Such a $mean~field$ mechanism, often referred to as isospin fractionation
\cite{baranPR}, is responsible for a reduction of the neutron
to proton ratio during the high density phase, with direct consequences
on particle production in inelastic $NN$ collisions.
On the other hand, 
the self-energy contributions, Eqs.\ref{selfen},  will influence the particle production at the
level of thresholds as well as of the phase space available in the final 
channel. As a matter of fact, the threshold effect is dominant and consequently the
results are nicely sensitive to the covariant structure of the isovector
fields, see fig.\ref{kaon1}.
The beam energy dependence of the $\pi^-/\pi^+$ (top) and  $K^0/K^+$ (bottom)
ratios is shown on the figure. 
At each beam energy we see an
increase of the $\pi^-/\pi^+$ and 
$K^{0}/K^{+}$ 
yield ratios with the models
$NL \rightarrow NL\rho \rightarrow NL\rho\delta$. 
The effect is larger for the $K^{0}/K^{+}$ compared to the $\pi^-/\pi^+$
ratio. This is due to the subthreshold production and to the fact that
the isospin effect enters twice in the two-step production of kaons, see
\cite{ferini06}. 
Interestingly the Iso-$EOS$ effect for pions is increasing at lower energies,
when approaching the production threshold.


\section{Hints from neutron stars}
Neutron stars appear as natural astrophysical laboratories to study the 
behavior of the EOS at high density. In particular, 
the structure and composition
of neutron stars is affected by the density dependence of the symmetry energy. 
Neutron stars are long-lived systems in $\beta$ equilibrium, hence in rather
different conditions with respect to the transient systems that are formed 
in nuclear reactions. The concept of EOS is very appropriate for these
objects, but experimental observations of their properties
are more difficult. 

It is generally believed that a neutron star (NS) is formed as a result of 
the gravitational collapse of a massive star in a type-II supernova. Hence a 
protoneutron star (PNS) appears, a very hot and lepton-rich object, where
neutrinos are temporarily trapped. The following evolution of the PNS is 
dominated by neutrino diffusion, which eventually results in cooling. 
The star stabilizes at practically zero temperature, and no trapped neutrinos
are left. 
\begin{figure}
\begin{center}
\includegraphics[scale=0.36,angle=-90]{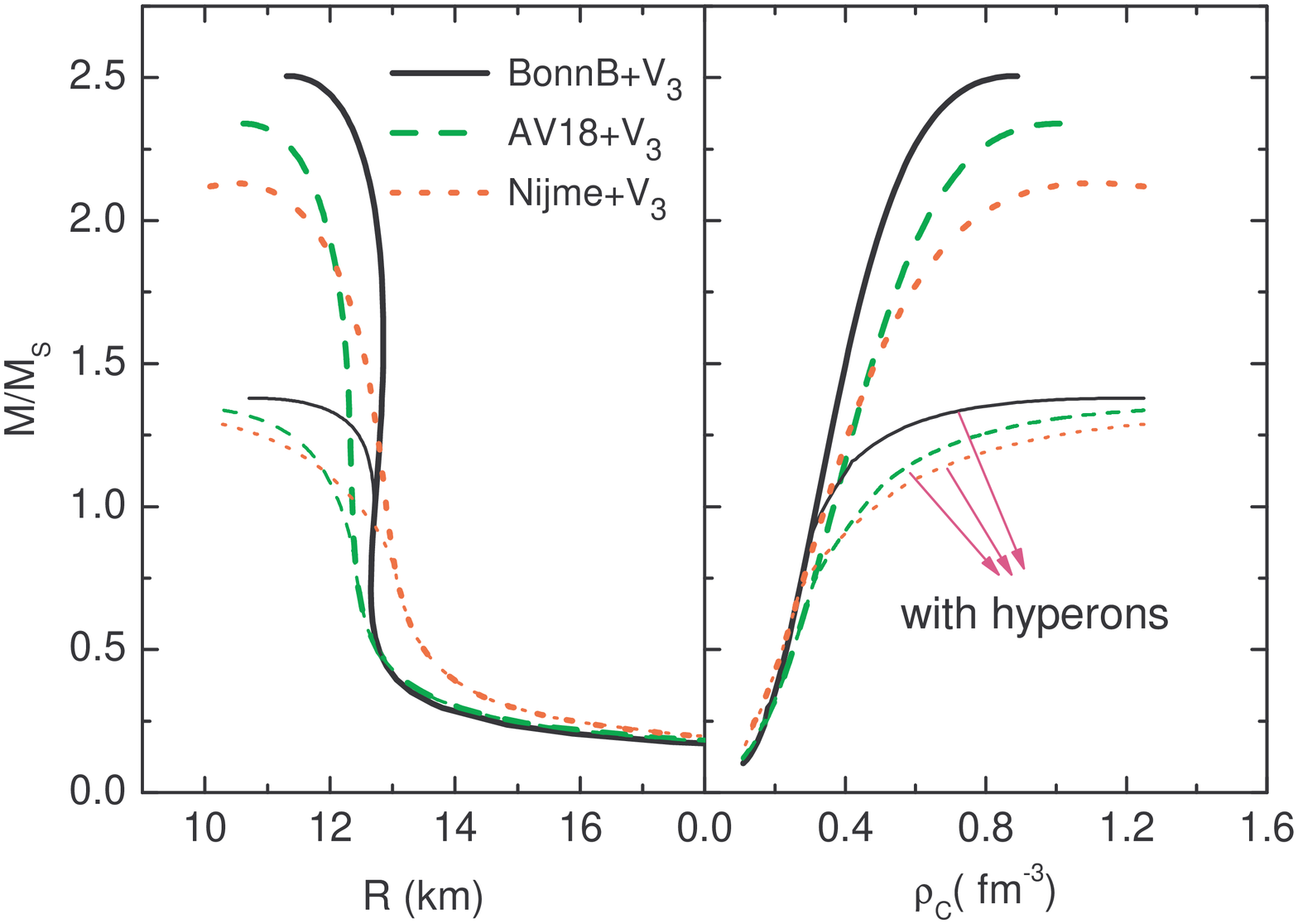}
\caption{
Mass-radius (left panel) and mass-central density (right panel) relations
of neutron stars evaluated with different nucleonic EOS. Taken from Ref.\cite{Lomb08}.
}
\label{hyper}
\end{center}
\end{figure}
\begin{figure}[h]
\centering
\begin{minipage}{14pc}
\includegraphics[width=14pc]{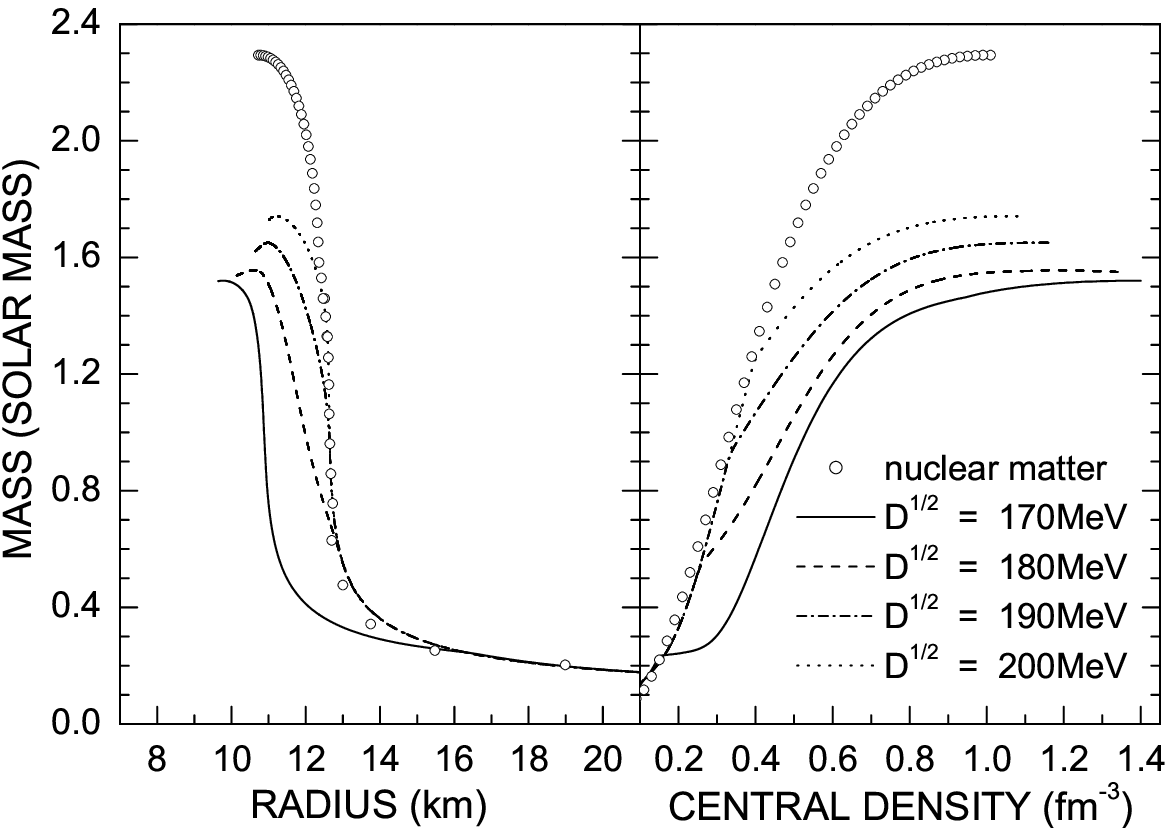}
\caption{\label{fio}
Mass-radius (left panel) and mass-central density (right panel) relations
of hybrid stars for four values of the confinement parameter
$D$. The nucleonic EOS is the one corresponding to the BOB interaction, see fig.2.  
Taken from Ref.\cite{Lomb08_2}.
}
\end{minipage}\hspace{2pc}%
\begin{minipage}{16pc}
\includegraphics[angle=+270,width=16pc]{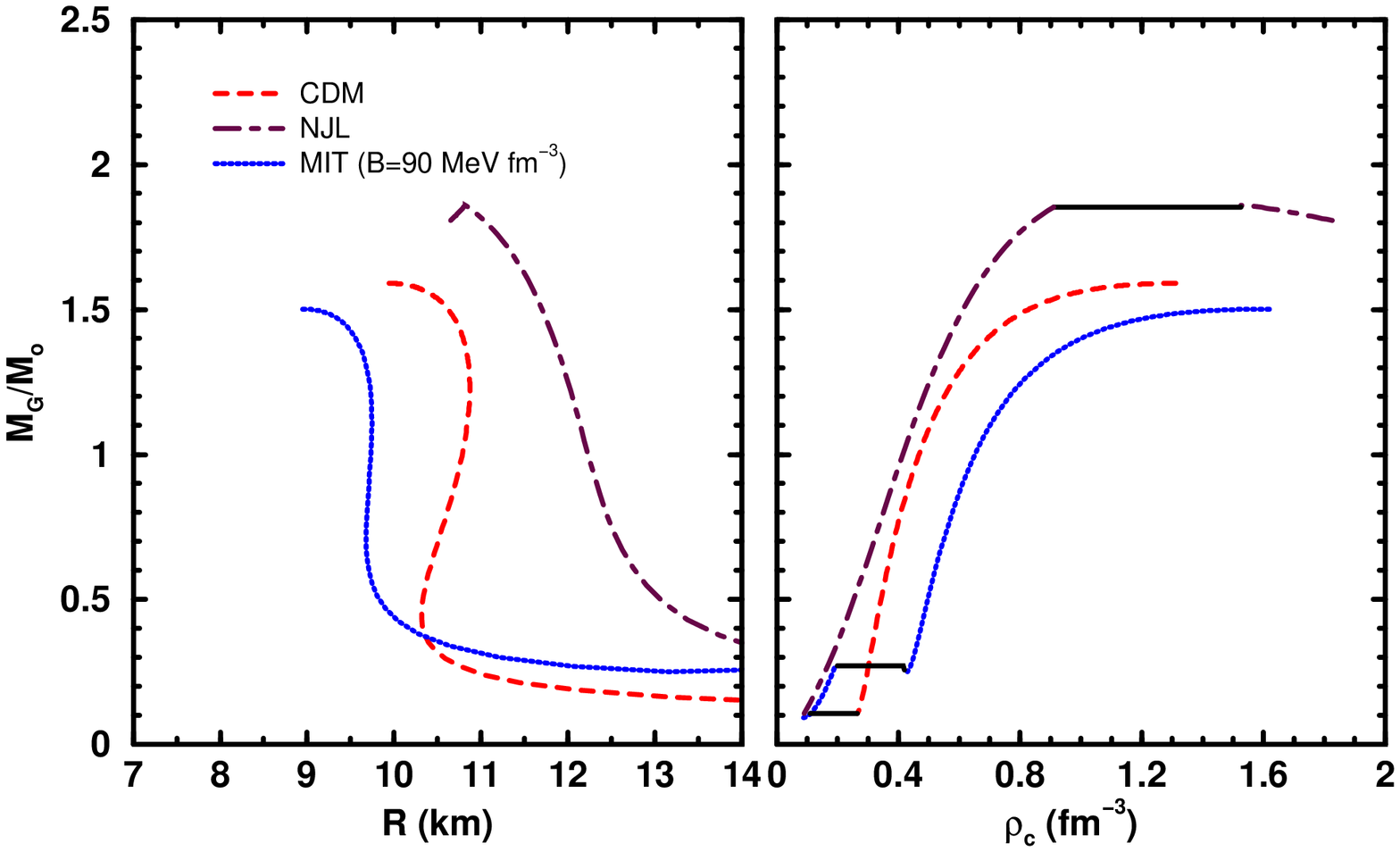}
\caption{\label{fio1}
Mass-radius (left panel) and mass-central density (right panel) relations
of hybrid stars, corresponding to three different models of the quark 
phase. The hadronic EOS includes hyperons. Adapted from Refs.\cite{Fio08},\cite{Fio08_1}.
}
\end{minipage} 
\end{figure}

The dynamical transformation of a PNS into a NS could be strongly influenced by
a phase transition in the central region of the star. In fact, the central 
particle density of a massive PNS may reach values larger than $1/fm^3$. 
In this density range the nucleon cores start to touch each other, and it is 
likely that quark degrees of freedom will play a role.     

For stars in which the strongly interacting particles are only baryons, 
the composition at given baryon density $\rho_B$  is determined by the 
requirements of $\beta$ equilibrium and charge neutrality,
involving nucleons, hyperons, leptons (electrons, muons) and neutrinos 
\cite{Lomb08_2,Fio08}.
These conditions fix, at the density $\rho_B$, and for given muon and 
electron numbers, the relative densities of 
all species. 
From the knowledge of these densities, using the Fermi-gas model for leptons,
(non-interacting) hyperons and for neutrinos and the usual 
thermodynamical relation
$ P = \rho^2 {{\partial(f/\rho)} \over {\partial \rho}} $ (being 
$f$ the
free-energy density) for neutrons and protons, one finally obtains the EOS of the 
matter that composes the star,  i.e. the pressure corresponding to $\rho_B$, $P(\rho_B)$. 
The result clearly depends on the nucleonic EOS adopted, usually taken from
microscopic calculations, see Section 2. 
Nucleon-hyperon interactions can also be included \cite{Fio08,Lomb08}.

To investigate the possible phase transition to quark matter in neutron stars, 
it is essential to know also the EOS of the quark matter. One usually deals with ordinary strange
quark matter (SQM) in $\beta$ equilibrium.  
The special problem in studying the EOS of ordinary quark 
matter is to treat quark confinement in a proper way. The conventional standard
approach is to add an extra constant term to the quark energy, the bag constant
$B$ (MIT bag model), that leads to a negative contribution to the pressure, to confine quarks in 
a finite volume \cite{Fio08}. 
Alternatively, in Ref.\cite{Lomb08_2} the quark confinement is treated
using the chirally dependent quark mass scaling, where the quark mass scales with 
density.
From the weak-equilibrium conditions of SQM, for given baryon and 
charge densities, it is possible to get the the relative densities of quarks (u,d,s) and electrons and
finally 
derive the EOS of the quark matter \cite{Lomb08_2}.   

The possible co-existence between quark and hadronic matter is studied imposing the
Gibbs conditions \cite{Landaustat}. 
Requiring the equality of the chemical potentials in the two phases, only two chemical potentials remain independent, 
such as $(\mu_n,\mu_p)$ for instance.  Their corresponding values 
as well as the quark
fraction, are determined by imposing to have the same pressure in the two phases
at the given total $\rho_B$ and electric charge.

Calculations show that the transition from the hadron to the mixed phase occurs at a density a bit less
than $0.15~fm^{-3}$, close to the saturation density. Of course, 
this is not the case 
in terrestrial laboratories, since the nuclear matter so far realized in exotic nuclei or heavy
ion collisions is much less neutron-rich than neutron stars.  However, the transition to the pure quark phase
occurs at much higher density, of the order of $0.6~fm^{-3}$. 
The transition density depends on the MIT bag parameter or, in the case of the density 
dependent quark masses, on the parameter usually named $D$ \cite{Lomb08_2}. 
One can note that,  increasing the bag parameter $B$, the pressure associated with the
quark phase decreases and the transition 
starts at higher density. 
It is worth mentioning that
the asymmetry of the quark phase is higher than the asymmetry of the 
co-existing hadronic phase.   

With the EOS which has the mixed and/or quark phases at hand, the structure of
hybrid stars can be studied by solving the Tolmann-Opennheimer-Volkoff (TOV) 
equation:
\begin{equation}
{dP \over dr} = -{GmE \over r^2}{(1 + P/E)(1 + 4\pi r^3 P/m) \over 1-2Gm/r},
\end{equation}
where $G$ is the gravitational constant, $r$ is the distance from the center of the star,
$E = E(r)$ and $P=P(r)$ are the energy density and pressure at the radius $r$, respectively. 
The EOS $P= P(E)$ is essentially the input of the equation. 
The subsidiary condition is:
\begin{equation}
dm/dr = 4\pi r^2 E,
\end{equation}
where $m(r)$ denotes the gravitational mass of the star. 
Starting with a central mass density $E_C$, Eq.(10) is integrated out until the pressure on the surface
equals the one corresponding to the density of iron. This gives the stellar radius $R$ and the gravitational
mass $M_G$.  In particular, one can investigate the maximum radius of stable stars. 

As one can see from fig.\ref{hyper},  
a remarquable reduction of the large range of the maximum mass obtained with the nucleonic EOS's
(1.8-2.5 $M_\odot$) to the nearly unique value of 1.3-1.4 $M_\odot$ is observed when only hyperons are included in
the calculation of the nuclear EOS, i.e. excluding the possible transition to the quark matter, but
implementing the nucleon-hyperon interaction in the EOS. 
The hyperons provide a self-regulating softening effect on the EOS: The stiffer the original nucleonic EOS,
the more it is softened by the earlier appearance at higher concentration of hyperons. 
Due to the softening of the EOS, the maximum stellar mass decreases. 

In any case, 
the maximum masses of hadronic (hyperonic) neutron stars are much too low in order to cover all
present observational values. 
Within the present descriptions, 
the introduction of nonhadronic ``quark'' matter is thus necessary
and heavier
neutron stars can only be hybrid stars. 
Some calculations including the possibility of a transition (or co-existence) to the quark phase
are shown in figs.\ref{fio},\ref{fio1}. 
Results obtained with different possibilities to model the quark phase and different nucleonic 
EOS are presented. 
With the inclusion of quarks, one obtains a maximum mass in the range 1.4-1.8 $M_\odot$, that depends 
on the model considered, but is 
closer to the experimental observations. 
The largest value is obtained adopting the Nambu-Jona Lasinio (NJL) model for the quark 
phase \cite{Fio08_1}.  Indeed, in this case,  the co-existence with the quark phase starts
at rather high density and no transition at all is observed if hyperons are also included. Moreover, 
the onset of the pure quark phase at the center of the PNS as the mass increases marks an instability 
of the star, i.e. the PNS collapses to a black hole at the transition point since the quark EOS is
unable to sustain the increasing central pressure due to gravity.

From the body of these calculations, one can conclude that 
limiting masses for PNS smaller than 2 solar masses are generally 
confirmed in all models .  

All that shows clearly the great relevance of observations on NS to our knowledge of the
high density nuclear EOS. The astrophysical observations could be able to rule out definite
EOS or put constraints on them.  However, the results discussed 
above indicate a larger sensitivity 
to the model adopted for the quark phase than to the nucleonic EOS. 

\section{On the transition to a mixed hadron-quark phase in HIC at relativistic energies}
The possibility of the transition to a mixed hadron-quark phase, 
at high baryon and isospin density, has been suggested also for the
dense and hot matter that is formed in HIC of charge asymmetric systems
at relativistic energies, such as for instance 
$Au + Au$ or $^{238}U+^{238}U$ (average proton fraction $Z/A=0.39$) at
$1~AGeV$.
\begin{figure}[h]
\centering
\includegraphics[width=16pc]{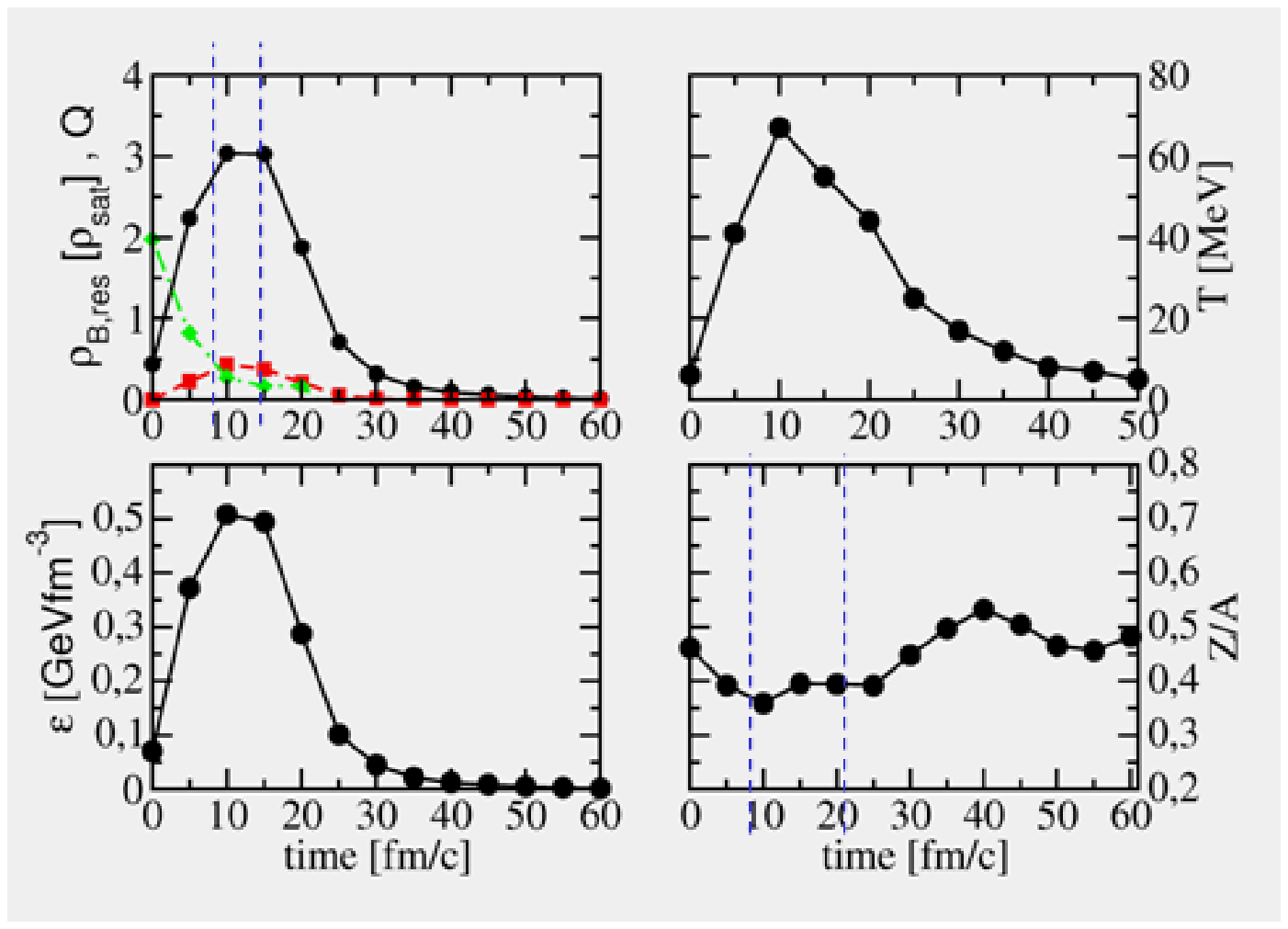}
\hskip 0.5cm
\includegraphics[angle=+90,width=16pc]{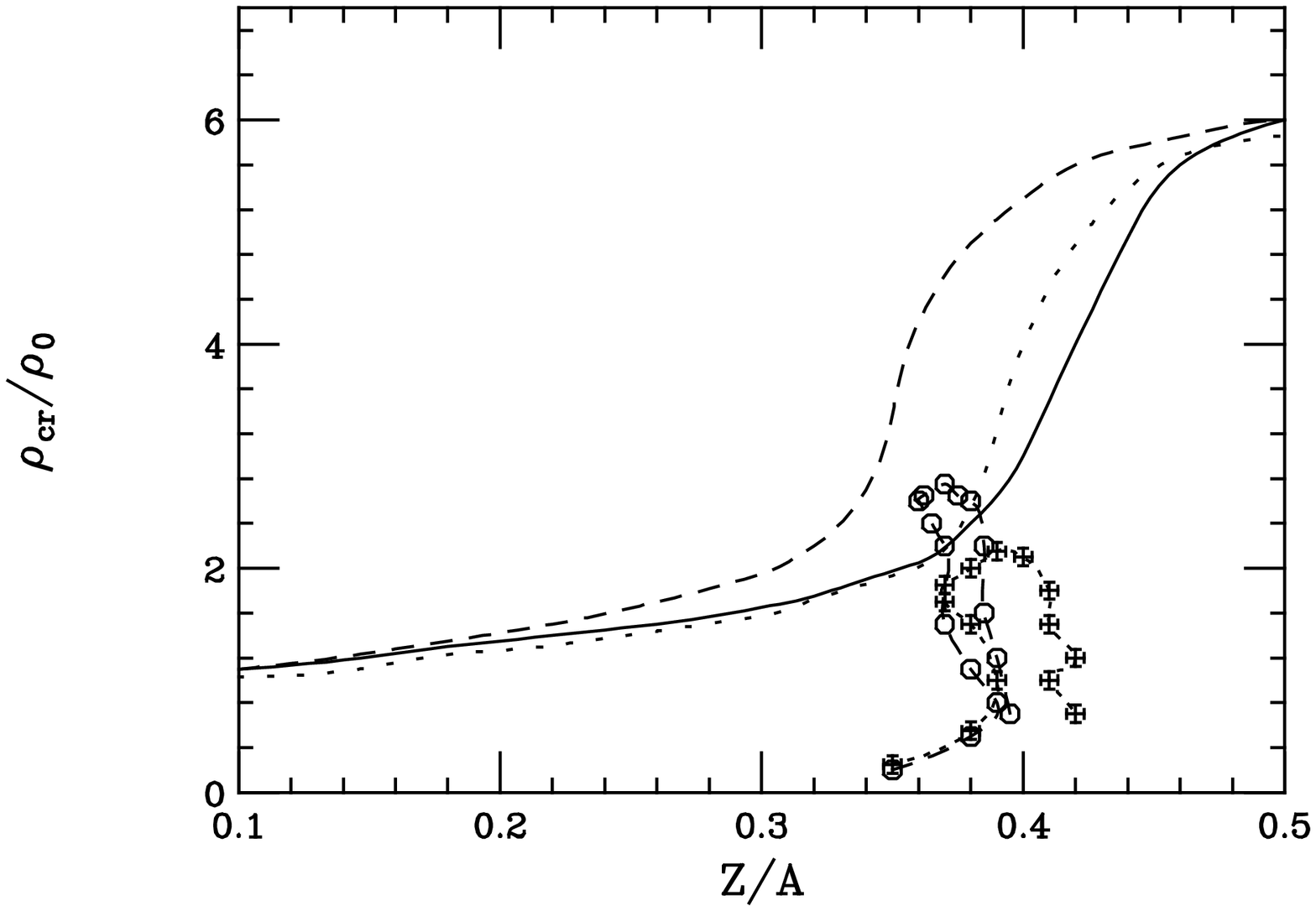}
\caption{\label{quark}
Left: Time evolution of density, temperature, energy density and Z/A
content reached in a central region of the reaction Au + Au at 1 AGeV, 
as obtained in RBUU calculations. 
Right: Variation of the transition density with proton fraction for various
hadronic $EOS$ parameterizations. Dotted line: $GM3$ $RMF$-model
\cite{GlendenningPRL18};
 dashed line: $NL\rho$ ; solid line: $NL\rho\delta$ . 
For the quark $EOS$: $MIT$ bag model with
$B^{1/4}$=150 $MeV$.
The points represent the path followed
in the interaction zone during a semi-central $^{132}$Sn+$^{132}$Sn
collision at $1~AGeV$ (circles) and at $300~AMeV$ (crosses).
From Ref.\cite{ditoro_dec}. 
}
\end{figure}

 
In  fig.~\ref{quark} (left) we discuss the evolution of temperature, 
baryon density, energy density and $Z/A$ in a space cell located in the c.m. of the system.
We note that a rather exotic nuclear matter is formed in a transient
time of the order of $10~fm/c$, with baryon density around $3-4\rho_0$,
temperature $50-60~MeV$, energy density $500~MeV~fm^{-3}$ and proton
fraction between $0.35$ and $0.40$, likely inside the estimated mixed 
phase region.

In fact 
the transition densities are largely isospin dependent
and, as seen in the previous Section
in the case of neutron stars, reach rather low values in very
asymmetric matter.

\begin{figure}[t] 
\centering
\includegraphics[angle=-90,width=8.0cm]{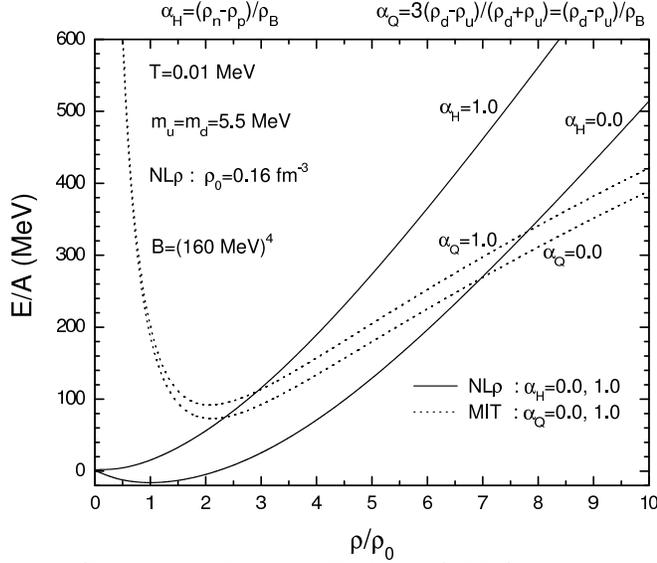} 
\vskip 1.0cm
\caption{$EOS$ of Symmetric/Neutron Matter: Hadron ($NL \rho$), solid lines,
vs. Quark (MIT-Bag), dashed lines. $\alpha_{H,Q}$ represent the isospin 
asymmetry parameters respectively of the hadron,quark matter:
$\alpha_{H,Q}=0$, Symmetric Matter; $\alpha_{H,Q}=1$, Neutron Matter.
Taken from Ref.\cite{Erice08}.
}
\label{isoparton} 
\end{figure} 

In the hadronic phase the charge chemical potential is given by
$
\mu_3 = 2 E_{sym}(\rho_B) \frac{\rho_3}{\rho_B}\, .
$ 
Thus, we expect critical densities
rather sensitive to the isovector channel in the hadronic $EOS$.

In fig.~\ref{quark}  (right) we show the crossing
density $\rho_{cr}$ separating nuclear matter from the quark-nucleon
mixed phase, as a function of the proton fraction $Z/A$, for different 
parameterizations of $E_{sym}$, see Section 5.2. 
The MIT bag model is adopted for the quark phase. 
We can see the effect of the
$\delta$-coupling towards an $earlier$ crossing due to the larger
symmetry repulsion at high baryon densities.
In the same figure  the paths in the $(\rho,Z/A)$
plane followed in the c.m. region during the collision of the n-rich
 $^{132}$Sn+$^{132}$Sn system, at different beam energies, are also reported. 
One can see that already at
$300~AMeV$ the systems may be on the border of the mixed phase, and 
well inside it at $1~AGeV$ \cite{ditoro_dec}. 

As a signature of co-existence with quark matter in HIC, 
one can expect a {\it neutron trapping}
effect, supported by statistical fluctuations as well as by a 
symmetry energy difference in the
two phases.
In fact while in the hadron phase we have a large neutron
potential repulsion (in particular in the $NL\rho\delta$ case), in the
quark phase we only have the much smaller kinetic contribution.
Observables related to such neutron ``trapping'' could be an
inversion in the trend of the formation of neutron rich fragments
and/or of the $\pi^-/\pi^+$, $K^0/K^+$ yield ratios for reaction
products coming from high density regions, i.e. with large transverse
momenta.  

From the above discussion and from the results reported in Section 6, 
it is clear that the possibility of a transition
to the quark phase crucially depends on the way to describe the quark EOS.
From this point of view,  
it appears extremely important to investigate the possible inclusion of  the 
Isospin degree of freedom in the effective approaches to the QCD 
dynamics \cite{Erice08}.
In  fig.\ref{isoparton} we report the EOS of symmetric and neutron matter
for hadronic (full lines) and quark (dashed lines) phases.  
One can see that the transition to the quark phase is 
particularly convenient because, 
in the quark models considered so far, 
the symmetry energy of the quark matter
takes contribution only from the kinetic energy and it is rather small.
In fact, the two dashed lines in fig.16 are very close, while in the hadronic
case the potential part of the symmetry energy is large.
     
Of course, these considerations are relevant also for the description
of the properties of neutron stars.

\vskip 0.5cm

\vskip -1.0cm
\section{Perspectives}

We have reviewed some aspects of the rich phenomenology associated with nuclear 
reactions and astrophysical observations, from which interesting hints
are emerging to constrain the nuclear EOS and, in particular, the largely 
debated density behavior of the symmetry energy.
The greatest theoretical uncertainties concerns the high density 
domain, 
that has the largest impact on the
understanding of the properties of neutron stars. 
This regime can be explored in terrestrial laboratories by using relativistic
heavy ion collisions of charge asymmetric nuclei. Differential collective flows
and meson production are promising observables.  
On the other hand, the behavior of the symmetry energy at low density 
can be accessed in reactions from low to intermediate energies, 
where collective excitations and fragmentation mechanisms are dominant.
A considerable amount of work has already been done in this domain.  
In the near future, thanks to the availability of both stable and rare
isotope beams, more selective analyses, also based on new exclusive 
observables, 
are expected to provide further stringent constraints. 

\vskip 1.0cm

\noindent
\section*{References}
\vskip 0.5cm


                                    %



\end{document}